# An Intuitive Approach to Special and General Relativity

Charles Francis

**Abstract:**
The *k*-calculus was advanced by Hermann Bondi as a means of explaining special relativity using only simple algebra (**Bondi H.:** *Relativity and Common Sense*, London, Heinemann, 1964). As used by Bondi, *k* is Doppler shift. This paper extends the *k*-calculus to include gravitational red shift and to develop techniques for an introductory treatment of general relativity in which the emphasis is on mathematical deduction from physical measurement procedure. Using ideas of geometric optics, geodesic motion is understood from the refraction of the wave function due curvature. The *k*-calculus gives a very simple derivation of Schwarzschild, showing that the geometry is equivalent to the existence of a fundamental minimum time, proportional to rest mass, between the interactions of elementary particles. The Newtonian approximation is seen from direct application of red shift to the wave function. Finally differential geometry is introduced, showing that the *k*-calculus gives an equivalent treatment of general relativity up to and including the general form of Einstein's field equation.

Charles Francis
Clef Digital Systems Ltd.
Lluest, Neuaddlwyd
Lampeter
Ceredigion
SA48 7RG

# An Intuitive Approach to Special and General Relativity

## 1   Background

Hermann Bondi expressed the view that '*with our modern outlook and modern technology the Michelson-Morley experiment is a mere tautology"* [1], the reason being that reference frames in space-time require light for their definition. This is not quite true, in part because the photon could have (or acquire) a non-zero but immeasurably small mass, but principally because in quantum field theory the amplitude for the creation of a particle and its annihilation at any non-synchronous point is non-zero, even outside the light cone, so the speed of individual photons is not constrained. Nonetheless in relativistic quantum field theory the speed of observable effects is bounded, and it is legitimate to discuss a maximum theoretical speed of information.

If we are to measure the time and distance of an event spacially separated from ourselves, then information must travel between us and the event. If we know the speed of information transfer, we can easily determine the time and distance of the event. But speed is defined in terms of time and distance, which leads to a paradox. The 4-coordinate of an event must be known before we can talk of the speed of information coming from it, but the speed of information must be known to determine the 4-coordinate of an event. To resolve the paradox we must find something fundamental, and base everything else on it. If we do not accept instantaneous action at a distance, then we may say that there is always a maximum speed of information, which we can call $c$. It is tautologous to say that the maximum speed of information is the same (up to scaling) in all reference frames, because there is no reference frame which does not depend for its definition on the maximum speed of information. In practice light does travel at $c$, to the limits of experimental accuracy, and for the purpose of this paper light is the carrier of information.

**Idealisation**

In the $k$-calculus the radar method is used to measure of time and distance coordinates, as describe in section 2, *Minkowski Coordinates*. This will be taken as the definition of space-time coordinates since any other method of measurement can be calibrated to give an identical result. Radar is preferred to a ruler, because it applies directly to both large and small distances, and because a single measurement can be used for both time and space coordinates.

What is discussed in this paper is an idealisation of radar. It is imagined that a radar pulse can be sent in any given direction at a precise time and that a reflected signal returns after a interval which can be precisely timed. There is no such thing as a perfectly confined wave packet, but, as previously remarked, the definition of time and space coordinates depends on the maximum theoretical speed of information in any direction, not on practical issues of signalling with e.m radiation. For example the simplest possible antenna, the dipole antenna, has a transmission/receiving pattern that resembles a figure 8. The transmitting pattern introduces an uncertainty in the direction in which photons are transmitted, and there is a corresponding uncertainty in our ability to determine the direction from which a received photon came. The smaller the object we are trying to detect by radar, the more its radiation pattern resembles that of a dipole. Hence, there is a relationship between the uncertainty in the position of an object and the size of the object compared to the wavelength of the probing signal. Indeed a radar pulse can be thought of as a wave packet describing the uncertainty in time of transmission of a photon.

To reduce uncertainty to achieve perfect eigenstates of position would require radar signals of infinitesimal wavelength and infinite energy. It is legitimate to discuss such idealisations since the definition of a metric does not depend on practical issues but on a bounding value. Similarly this paper will treat clocks and particles from a classical perspective, as though they could be used to define an exact origin of coordinates. This is a premise of the Copenhagen interpretation and it is recognised that uncertainty does not permit such an exact definition. But it is reasonable to apply the analysis to the limiting case of eigenstates of position and to use a space-time diagrams showing the reflection of a photon as sharply defined.

Quantum electrodynamics has shown that the exchange of photons is responsible for the electromagnetic force and radar ranging is an instance of photon exchange. Since electromagnetic forces are responsible for all the structures of matter in our macroscopic environment, it is not unreasonable to postulate that photon exchange is responsible for geometrical relationships internally within a body as well, by analogy with the process as it takes place in radar. This paper seeks to give a (semi)classical treatment and does not enter into such issues. A version of the arguments based on particle theoretic quantum electrodynamics is given in [2].

## 2   Minkowski Coordinates

There is room for confusion between two very similar questions, 'What is time?' and 'What is the time?'. The first question has something to do with consciousness, and our perception of time as a flow from past to future. It admits no easy answer, but is quite distinct from the second question and only the second question is relevant in the definition of space-time coordinates. The answer to the question 'What is the time?' is always something like 4:30 or 6:25.

**Definition:** The time is a number read from a clock.

There are many different types of clock, but every clock has two common elements, a repeating process and a counter. The rest of the mechanism converts the number of repetitions to conventional units of time. A good clock should provide accurate measurement and it should give a uniform measure of time. We cannot count less than one repetition of the process in the clock, so for good resolution the process must repeat as rapidly as possible. In a uniform clock, the repeating process must repeat each time identical to the last, uninfluenced by external matter. Since 1967 the standard second has been by definition 9,192,631,770 periods of the unperturbed microwave transition between the two hyperfine levels of the ground state of $Cs^{133}$. This process is chosen for the practical reasons that it is rapid and not easily perturbed by external matter, but it is an arbitrary choice. If in the future we are able to build better clocks using another oscillation, we will be free to change the definition, but a second will still be a count of unperturbed oscillations, not an absolute flow like Newtonian time.

Clocks may be called identical if they measure identical units of time when they are together, and if moving them does not noticeably affect the physical processes of their operation. A clock defines the time, but does so only at one place. A space-time reference frame also requires a definition of distance, and a definition of time at a distance from the clock. Both are provided for by the radar method.

**Definition:** The distance of an event is half the time for light to go from a clock to the event, to be reflected and to return to the clock. The time of reflection is the mean of time sent and time of return.

**Definition:** A reference frame is the set of possible results of time, distance and direction measurements based on a given clock. It does not need to be specified whether coordinates are polar or cartesian or other, since we may change coordinate system with a matrix transformation (see section 3, *Vectors*), and for convenience a reference frame is described as the set of coordinate systems defined from a given clock using radar.

**Definition:** In Minkowski coordinates a reference frame has one time and three cartesian space axes.

To use radar we must know the speed of light (if distance were defined using a ruler, then to measure the time at an event we would still need to know the speed of a message from the event). But now we have a paradox. To measure speed we conduct a time trial over a measured distance, but first time must be defined at both ends of the ruler, which requires knowledge of the speed of light. We know no other way to measure the time of an event at a distance from a clock; if we synchronise two clocks by bringing them together, we have no guarantee that they remain synchronised when they are separated, unless light is used to test their synchronisation. Thus the speed of light is an absolute constant because measurement of speed requires a reference frame, which requires light for its definition. An experiment to determine the speed of light actually measures the conversion factor from natural units in which the speed of light is 1.

Space-time diagrams are defined such that lines of equal time are horizontal and lines of equal distance are vertical, and such that light is drawn at $45^o$ (figure 1). The radar method also measures direction. It will be seen that the algebra is formally identical for 3-vectors in cartesian coordinates. Indeed space-time diagrams are best understood as showing the radial coordinate in 3 dimensions and do not give only a one dimensional treatment. The radar method defines distance in units of time; this paper will use natural units in which the speed of light is 1, and speed is measured as a fraction of the speed of light. To restore conventional units substitute $v \to v/c$ Radar is preferred to a ruler, because it applies directly to both large and small distances, and because a single procedure can be used for both time and space coordinates.

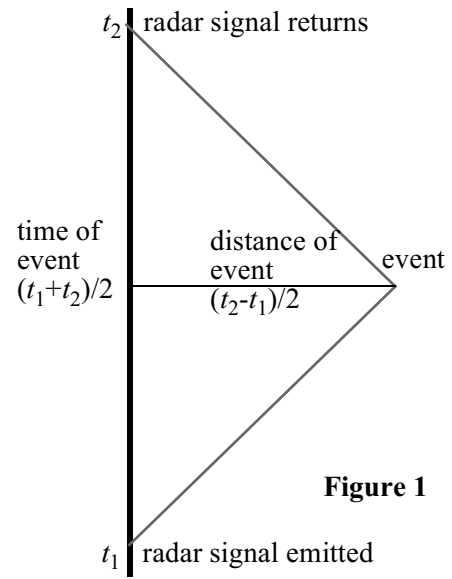

**Figure 1**

If we wish to compare our reference frame with the reference frame of a moving observer, we need to know what unit of time the moving observer is using. By definition two clocks give the same unit of time if they are calibrated to Cs$^{133}$. Figure 2 shows the reference frame defined by an observer in a moving space craft, as it appears to us, and our reference frame as it appears to him. The moving observer represents himself with a vertical axis, and he would draw us at an angle. In his diagram our reference frame appears distorted.

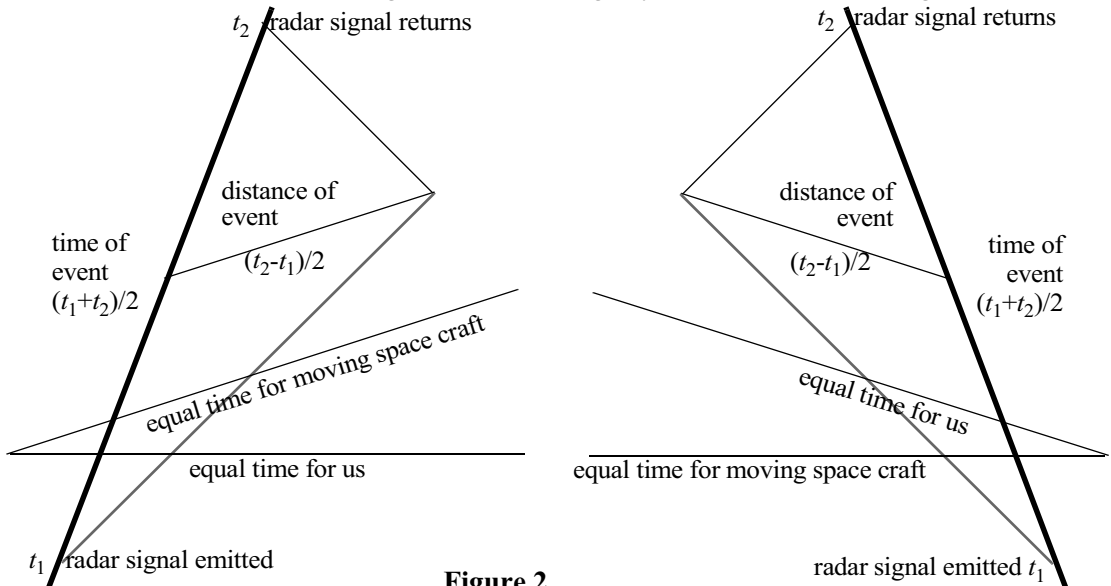

**Figure 2**

Uniform motion means motion shown by a straight line on a space-time diagram. In figure 3, a space craft is uniformly moving in the Earth's reference frame. The space craft and the Earth have identical clocks and communicate with each other by radio or light. The Earth sends the space craft two signals at an interval $t$. The space craft receives them at an interval $kt$ on the space craft's clock. $k \in \mathbb{R}$ is immediately recognisable as red shift (by considering the signals as the start and stop of a burst of light of a set number of wavelengths of a set frequency). Similarly if the observer on the space craft sends two signals at an interval $t$ on his clock, they are received at an interval $k't$ on the Earth. To analyse the relationship between $k$ and $k'$ we must first ensure that no external matter affects the motion of either space-craft or Earth, by defining

**Definition:** An *inertial* object is one whose motion is not affected by the direct action of other objects (to the limits of experimental accuracy). An inertial reference frame is one such that inertial matter initially at rest near the origin will remain at rest.

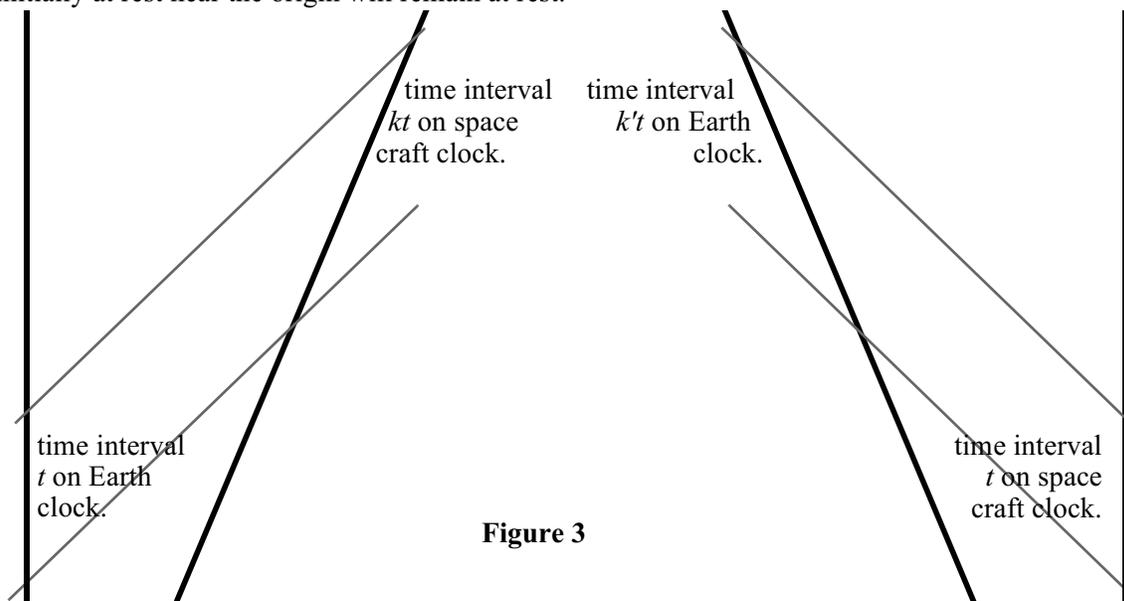

**Figure 3**

Provided only that they are inertial, there is no fundamental difference between the matter in the space craft and the matter in the Earth. The space craft can be regarded as stationary, and the Earth as moving. This implies that signals sent by the space craft to the Earth are also subject to red shift. The defining condition for the special theory of relativity is that we use reference frames such that

**Definition:** In Minkowski space-time red shift between inertial frames is both constant and equal for both observers, $k = k'$.

This condition is not universally true, even for inertial observers, as is described in the general theory of relativity. But if there is no action on a body it has no preferred orientation in space-time. Then for inertial frames $k = k'$ whenever their origins coincide. So we expect that Minkowski space-time applies as a local approximation everywhere. For the remainder of this section it is assumed that $k = k'$, this being the condition for the special, rather than the general, theory of relativity.

**Theorem:** (Time dilation, figure 4) The time $T$ measured by a space craft's clock during an interval $t$ on the Earths clock is given by

$$T = t\sqrt{1 - v^2} \qquad 2.1$$

*Proof:* The space craft and the Earth set both clocks to zero at the moment the space craft passes the Earth. The space craft is moving at speed $v$, so by definition, after time $t$ on the Earth clock, the space craft has travelled distance $vt$. Therefore Earth's signal was sent at time $t - vt$, and returned at time $t + vt$. For inertial reference frames, if the space craft sends the Earth signals at an interval $t$ the Earth receives them at an interval $kt$. So

$$T = k(t - vt). \qquad 2.2$$

Then by applying the Doppler shift again for the signal coming back

$$t + vt = k^2(t - vt) \qquad 2.3$$

Eliminating $k$ gives 2.1, the formula for time dilation.

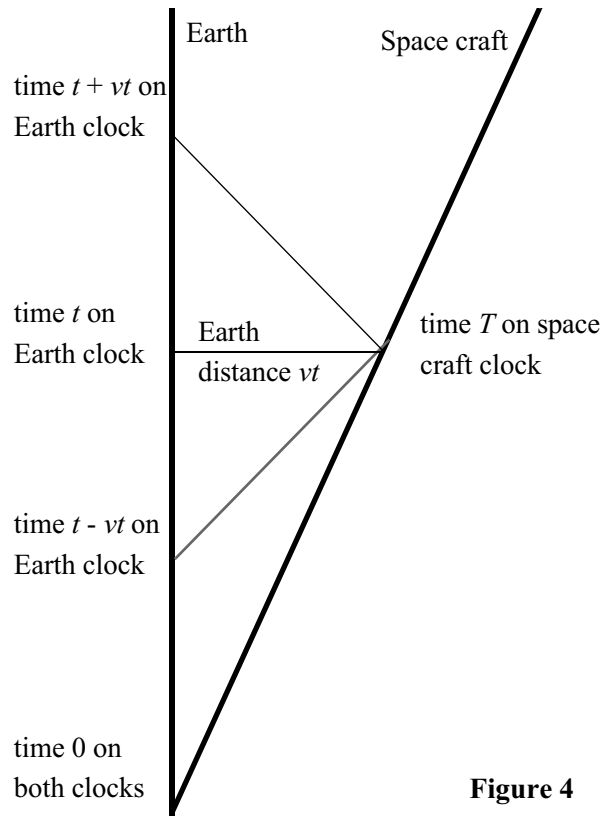

Figure 4

**Theorem:** (Lorentz Contraction, figure 5) A distance $d$ on the earth is measured on a space craft to be

$$D = \frac{d}{\sqrt{1 - v^2}} \qquad 2.4$$

*Proof:* The bow and stern of the space craft are shown as parallel lines. The space craft's clock is in the bow. The space craft and Earth set their clocks to zero when the bow passes the Earth clock. Earth uses radar to measure the distance, $d$, to the stern, by sending a signal at time $-d$, which returns at time $d$ on the Earth clock. The same signal is used to measure $D$ on the spaceship. By the Doppler shift, the outgoing signal passes the bow at time $-(d/k)$ on the space craft's clock, and the returning signal reaches the bow at time $kd$. So, according to the moving space craft

$$D = (kd + d/k)/2 \qquad 2.5$$

Eliminating $k$ using 2.3 gives 2.4.

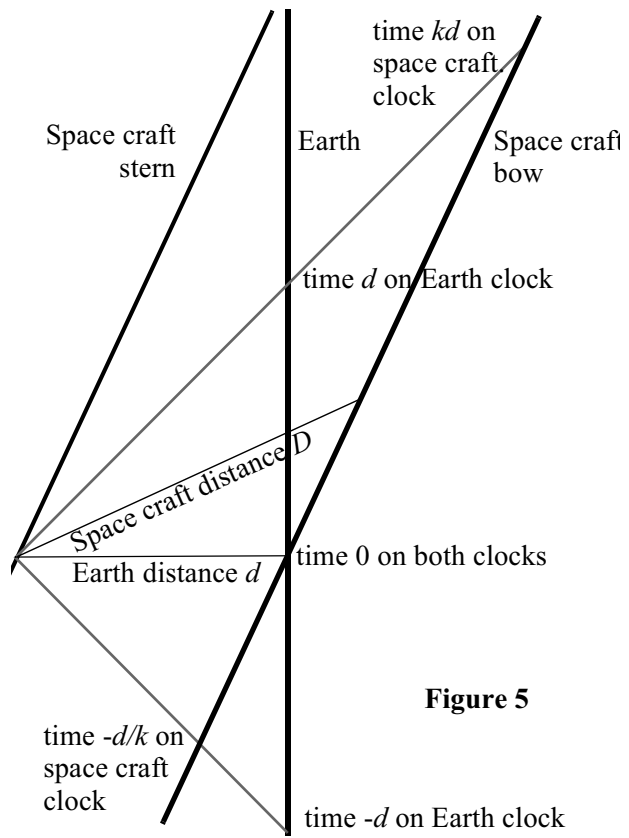

Figure 5

## 3   Vectors

**The Principle of General Relativity**

The principle of general relativity states *the laws of physics should be the same irrespective of the coordinate system which a particular observer uses to quantify them*. This is a form of the principle of homogeneity, that *the behaviour of matter is everywhere the same*. Laws which are the same in all coordinate systems are most easily expressed in terms of invariants, quantities which are the same in all coordinate systems. The simplest invariant is an ordinary number or scalar. Another invariant, familiar from classical mechanics, is the vector. A change of coordinates has no effect on a vector, but it changes the description of a vector in a coordinate system. Vector transformation laws are found by defining displacement vectors in Minkowski space, and used to define more general vector and tensor quantities which are the same in all coordinate systems. Then the form of the principle of homogeneity most directly applicable in relativity is the principle of general covariance, *The equations of physics have tensorial form*.

**Definition:** A space-time vector is the difference in the coordinates of two events. When no ambiguity arises space-time vectors are simply called vectors. Conventionally a space-time vector is written

$$x = (x^0, x^1, x^2, x^3) \qquad 3.1$$

where $x^0 = t$ is the difference in time coordinates of the events, known as the time component, and $\boldsymbol{x} = (x^1, x^2, x^3)$ is an ordinary displacement vector in 3 dimensions, or 3-vector.

**Velocity 4-Vector and Proper Time**

For any body a displacement vector can be defined as the difference between its co-ordinates at two different times. The 3-velocity of the body in a given coordinate system is then the space-coordinates divided by the time coordinate. There is a particular coordinate system in which the body is at rest, so that the displacement vector reduces to $(\tau, 0, 0, 0)$, where $\tau$ is time measured by a clock at rest with respect to the body, and is known as proper time. A velocity 4-vector is defined by differentiating with respect to proper time. The velocity 4-vector is $(1, 0, 0, 0)$ in a frame in which the body is at rest.

**$E=mc^2$**

**Definition:** Since momentum $\boldsymbol{p} = (p^1, p^2, p^3)$ is conserved in Newtonian mechanics, it is a vector with central importance. It is natural to define a corresponding 4 vector $p = (p^0, p^1, p^2, p^3)$

**Theorem:** (Mass shell condition) $p^0$ can be identified with energy $p^0 = E$ such that if $m$ is rest mass

$$m^2 = E^2 - \boldsymbol{p}^2 \qquad 3.2$$

**Corollary:** In conventional units ($c \neq 1$) 3.2 takes the form $m^2 c^4 = E^2 - \boldsymbol{p}^2 c^2$, and for a body which is not moving $\boldsymbol{p} = \boldsymbol{0}$ so this reduces to $E = mc^2$

*Proof:* There is a particular coordinate system in which $p$ represents a state of rest. In this coordinate system there is some number $\mu$ such that $p = (\mu, \boldsymbol{0})$ Then in the frame of an observer moving at velocity $\boldsymbol{v}$ relative to the clock $p$ has coordinates given by the formulae for time dilation, 2.1 and Fitzgerald contraction, 2.4

$$p = (\mu(1 - v^2)^{-1/2}, \mu\boldsymbol{v}(1 - v^2)^{-1/2}) \qquad 3.3$$

Since the 3-vector part $\boldsymbol{p} = \mu\boldsymbol{v}(1 - v^2)^{-1/2}$ must reduce to the Newtonian formula $\boldsymbol{p} = m\boldsymbol{v}$ for non relativistic velocities $|v| \ll 1$ we identify $\mu = m$, known as rest mass. The time component of 3.3 is

$$p^0 = mv(1-v^2)^{-1/2} \approx m + \tfrac{1}{2}mv^2 \qquad 3.4$$

$\tfrac{1}{2}mv^2$ is the classical expression for kinetic energy and Einstein removed the arbitrary constant in Newtonian energy by defining energy to be $E = p^0$. 3.2 and the corollary $E = mc^2$ follow at once.

**Coordinate Transformation**

General coordinate transformation incorporating rotation, shear, stretch, and boost, 3.3, from unprimed axes to new axes denoted prime can be described by multiplying any vector $x$ by a matrix

$$x^{\mu'} = \sum_\nu k^{\mu'}_\nu x^\nu \qquad 3.5$$

In general the matrix $k^{\mu'}_\nu$ is not constant but may have different components at each point in space-time. This will be the case for curvilinear coordinates, and it will inevitably be the case in curved space-time.

**Einstein Summation Convention.** It turns out that whenever an index is repeated above and below it is summed over. This is taken as implicit, so it is not necessary to write $\Sigma$ in expressions like 3.5, which reduces to

$$x^{\mu'} = k^{\mu'}_\nu x^\nu \qquad 3.6$$

**Definition:** The law of coordinate transformation, 3.6, between locally defined Minkowski co-ordinates is called Lorentz transformation.

Interchanging the axes and the suffixes we have

$$x^\nu = k^\nu_{\mu'} x^{\mu'} \qquad 3.7$$

and hence from 3.5

$$x^\nu = k^\nu_{\mu'} k^{\mu'}_\lambda x^\lambda \qquad 3.8$$

This is true for any vector $x$, so it follows that

$$k^\nu_{\mu'} k^{\mu'}_\lambda = \delta^\nu_\lambda \qquad 3.9$$

where the Kronecker delta $\delta^\nu_\lambda = 1$ if $\lambda = \mu$ and $\delta^\nu_\lambda = 1$ otherwise.

**The Scalar Product**

**Definition:** If $x = (x^0, x^1, x^2, x^3)$ and $y = (y^0, y^1, y^2, y^3)$ are vectors in space-time expressed in Minkowski coordinates then the scalar product is

$$x \cdot y = x^0 y^0 - x^1 y^1 - x^2 y^2 - x^3 y^3 \qquad 3.10$$

**Theorem:** The scalar product is invariant under coordinate transformation

*Proof:* Straightforward from 3.9.

**Definition:** The invariant squared magnitude of a vector $x$ is given by the scalar product with itself

$$x^2 = x \cdot x \qquad 3.11$$

**Contravariant Vectors**

**Definition:** A contravariant vector, $a$, is an ordered 4-tuplet

$$a = (a^0, a^1, a^2, a^3) \qquad 3.12$$

which obeys the same transformation law (3.5) as a space-time vector under change of coordinates.

The definition of a space-time vector only holds in Minkowski space-time ($k = k'$), but the definition of a contravariant vector is made in terms of a coordinate transformation, and applies also to vectors acting at points in other geometries. This is adequate since it is always possible to define space-time vectors in a region of space which is small enough to be considered Minkowski up to the limit of experimental accuracy (this statement can be made precise in differential geometry by defining vectors in the limit as the size of the region goes to zero).

**Covariant Vectors**

When multiplying matrices it is conventional to multiply the terms of rows by the terms of columns and sum. But in the language of tensors used for relativity this notation is inadequate. Instead we adopt the convention that column matrices are represented with a superfix, like 3.12 and row matrices are represented with a suffix, and then we forget that we ever used to distinguish columns from rows. Instead of multiplying columns by rows term by term, we multiply terms with a suffix by those with a superfix. The new notation is more powerful and simplifies the definition of generalised matrices having more than one index at top and or bottom.

**Definition:** For any contravariant vector $x = (x^0, x^1, x^2, x^3)$ define a covariant vector

$$(x_0, x_1, x_2, x_3) = (x^0, -x^1, -x^2, -x^3) \qquad 3.13$$

Then (using the summation convention) 3.10 can be written in the form

$$x \cdot y = x_\mu y^\mu \qquad 3.14$$

**Theorem:** The transformation law for a covariant vector is

$$x_{\mu'} = k_{\mu'}^\nu x_\nu \qquad 3.15$$

*Proof:* Let $A^\mu$ be a contravariant vector. Then $A^\mu x_\mu$ is an invariant

$$A^{\mu'} x_{\mu'} = A^\nu x_\nu = k_{\mu'}^\nu A^{\mu'} x_\nu \qquad 3.16$$

This is true for any values of $A^{\mu'}$, so 3.15 follows immediately.

**The Metric**

It is convenient to define index raising and lowering matrices

$$g_{\mu\nu} = \begin{bmatrix} 1 & 0 & 0 & 0 \\ 0 & -1 & 0 & 0 \\ 0 & 0 & -1 & 0 \\ 0 & 0 & 0 & -1 \end{bmatrix} \text{ and } g^{\mu\nu} = \begin{bmatrix} 1 & 0 & 0 & 0 \\ 0 & -1 & 0 & 0 \\ 0 & 0 & -1 & 0 \\ 0 & 0 & 0 & -1 \end{bmatrix} \qquad 3.17$$

**Definition:** $g$ is the metric for Minkowski space-time. We have

$$x_\mu = g_{\mu\nu} x^\nu \text{ and } x^\mu = g^{\mu\nu} x_\nu \qquad 3.18$$

and the scalar product can be written

$$x \cdot y = g_{\mu\nu} x^\mu y^\nu = g^{\mu\nu} x_\mu y_\nu \qquad 3.19$$

In general the metric is a function of position, but since Minkowski space-time applies as a local approximation everywhere the Minkowski metric is valid at the origin of every inertial reference frame defined by the radar method. The Minkowski metric is often denoted by a constant $\eta_{\mu\nu}$ whereas $g_{\mu\nu} = g_{\mu\nu}(x)$ will denote a varying metric with the raising and lowering property, and used for the general definition of a scalar product in any coordinate system as seen in 3.19.

**Tensors**

Given any two any vectors $A^\mu$, $B^\nu$ we may form the sixteen quantities $A^\mu B^\nu$, which form the components of a second rank tensor, also called the outer product or tensor product of $A^\mu$ and $B^\nu$. Then in general a second rank tensor is any quantity which can be formed from a sum of tensor products.

$$T^{\mu\nu} = A^\mu B^\nu + A'^\mu B'^\nu + A''^\mu B''^\nu + \ldots \qquad 3.20$$

Tensors of any rank, i.e. with any number of contravariant and covariant indices can be formed in the same way, and by lowering indices with $g_{\mu\nu}$. Each index of a tensor obeys the same transformation law, 3.5 or 3.15 as for a vector. To preserve the scalar product under general coordinate transformation we require that for any vectors $A^\mu$, $B^\nu$

$$g_{\alpha'\beta'}A^{\alpha'}B^{\beta'} = g_{\mu\nu}A^\mu B^\nu = g_{\mu\nu}k^\mu_{\alpha'}k^\nu_{\beta'}A^{\alpha'}B^{\beta'} \qquad 3.21$$

And since this is true for all values of $A^{\alpha'}$, $B^{\beta'}$ we can infer

$$g_{\alpha'\beta'} = g_{\mu\nu}k^\mu_{\alpha'}k^\nu_{\beta'} \qquad 3.22$$

So the transformation law for $g_{\mu\nu}$ is identical to that of covariant vectors for each suffix independently. Similarly the transformation law for $g^{\mu\nu}$ is identical to that of contravariant vectors for each superfix independently. $g$ is therefore a tensor. $g$ is called the metric tensor. For other indexed quantities, non-tensors, the relationship between superfixes and suffixes using $g$ is maintained, but non-tensors are distinguished from tensors because their indices do not satisfy vector transformation laws.

## 4    Non-Euclidean Geometry

Newtonian space is rooted in Euclidean geometry, which, as metaphysics, has been repeatedly and severely challenged by philosophers and mathematicians, notably by Descartes and Leibniz. But alternative geometries were not seriously considered until the possibility had been suggested by Gauss that Euclid's fifth postulate, that parallel lines can extend indefinitely always the same distance apart might not be true of the real world.

**Intrinsic and Extrinsic Curvature**

For historical reasons non-Euclidean geometries are called curved. But this is not curvature in the familiar sense of a curved surface. We distinguish between intrinsic and extrinsic curvature. Extrinsic curvature is the familiar concept of curvature, the shape of a two dimensional surface as it appears in three dimensional space. Intrinsic or Gaussian curvature refers to the geometrical properties of a space found from measurements taken within the space, and is perhaps better understood as stretching, or distortion of a measured length scaled to the apparent length shown on a flat diagram such as figure 8.

Gaussian curvature was originally developed to analyse the properties of two dimensional curved surfaces embedded into three dimensions. These surfaces also have extrinsic, or ordinary curvature. This historical fact, related to Gauss' employment in cartography, led to the terminology "curvature" for non-Euclidean geometry. But, as the following examples show, intrinsic and extrinsic curvature are quite distinct ideas. Intrinsic curvature can apply to spaces which appear flat, just as extrinsic curvature can apply to spaces on which Euclidean geometry holds.

**The Geometry of a Curved Surface**

Gauss laid the foundation for Riemannian geometry by finding a way of describing a curved surface in terms of the distances between any two points along a given path in the surface. When you construct two equal lines, AD and BC, perpendicular to AB, using only shortest paths on a sphere you find that CD is less than AB, so the parallel postulate does not hold (figure 6) By the same token the circumference of a circle is less than $2\pi r$.

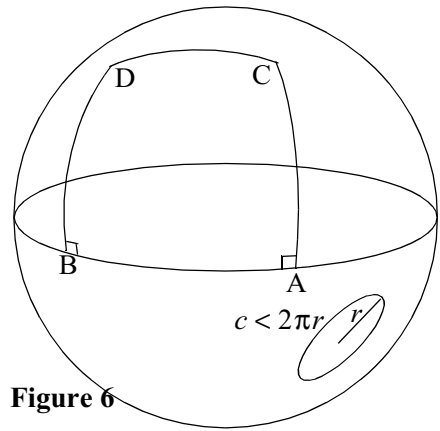

Figure 6

**Cylinder.** (figure 7) A cylinder has normal, extrinsic curvature, but no intrinsic, or Gaussian, curvature, because you can make a cylinder by rolling up a piece of paper without changing any of the distance relationships between points. The circumference of a circle on the paper is still $2\pi r$ when the paper is rolled up.

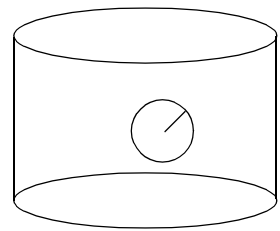

Figure 7

**World map.** (figure 8) A map of the world can be drawn on a flat piece of paper, but the distances referred to by the map are geographical. The map is flat by normal standards, but it has exactly the same geometry as the world itself, so it is the geometrical equivalent of a sphere.

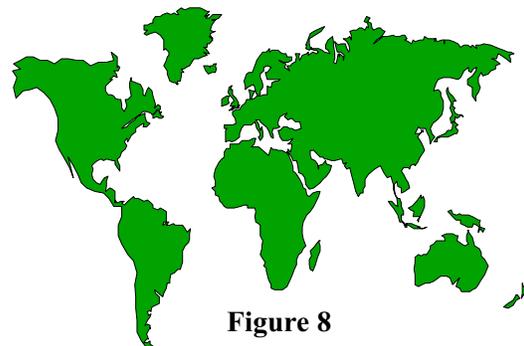

Figure 8

**Reflection in a Curved Mirror.** The image in a concave or convex mirror looks curved, so it is reasonable to say that it has extrinsic curvature — it is curved according to any normal definition. But the distance between two points in the reflection would perhaps be best measured by the reflection of a ruler, and is the same as the original. The reflection is, therefore, flat by the definition of Gaussian, or intrinsic, curvature. The same would apply to an image in a magnifying glass.

**Cone.** (figure 9) Like the cylinder you can make a cone by rolling up a flat piece of paper. It is flat everywhere except at the apex. A circle enclosing the apex has circumference less than $2\pi r$, a circle anywhere else has circumference equal to $2\pi r$.

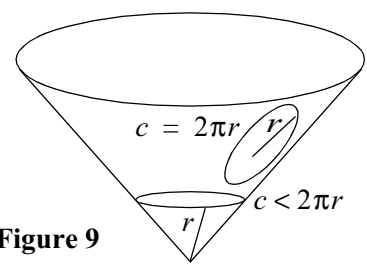

Figure 9

**Space-time.** The curvature of space-time is intrinsic, because it refers to geometrical properties defined by measurement within space-time. Extrinsic curvature makes no sense when applied to space-time, because there is no outside to look at it from.

**Positive and Negative Curvature**

Usually curved geometries are rigorously described by Riemann's theory of differential manifolds and the tensor calculus. In this treatment we will use an equivalent intuitive characterisation based on the idea that a map can be drawn using a non-constant scale factor. If we measure the distance AB between any two points, and measure two equal distances AD = BC = $h$ perpendicular to AB, as in figure 10, then we have no prior reason to assume that AB is equal to DC, as measured by an observer at D, and we have CD = $kd$ where $k$ is a non-constant scale factor. In general $k$ may have dependencies on both position and direction.

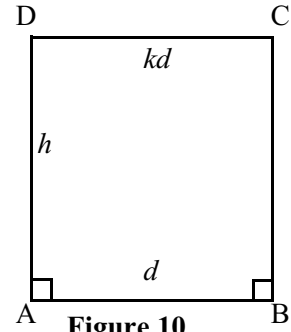
**Figure 10**

Gaussian curvature can be characterised by geometric properties. For example we can characterise curvature by comparison with the parallel postulate: In figure 10 we have CD = $k$AB

        Negative curvature:         $k$ increasing with $h$ = AD
        Zero curvature:             $k$ constant as $h$ increases
        Positive curvature:          $k$ decreasing as $h$ increases

The parallel postulate is a not an ideal criterion on which to base a categorisation of geometry, because geometrical systems start from a point, namely the origin of the coordinate system, not a line. An equivalent, and more natural, characterisation of a geometry is found by considering the length of an arc, CD, of a circle of radius, $s$, subtended by an angle, $\theta$, at the origin, O (figure 11). Use a very small angle, $\theta$, and drop perpendiculars of equal length from CD to a base line, AB, through the origin O. Then, in Euclidean geometry the length of CD is $r\theta$, almost equal to AB. But in general the length of CD is $kr\theta$, and the value of $k$ characterises the geometry according to the relationships

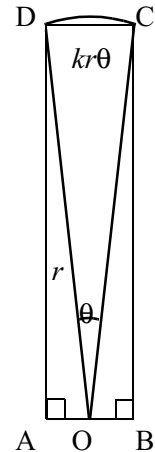

        Negative curvature:         $k$ increasing with $r$
        Zero curvature:             $k$ constant as $r$ increases
        Positive curvature:          $k$ decreasing as $r$ increases

**Figure 11**

This characterisation of positive and negative curvature conveys an idea of curvature, but it is simplistic. In figure 8 there are distortions both of scale and direction of the NS/EW coordinate system of the earth's surface compared to the Euclidean geometry of the paper and generally $k = k^{\mu}_{\nu'}(x)$ has dependencies on both position $x$ and the relative directions $\mu$ and $\nu'$ of the real and Euclidean axes.

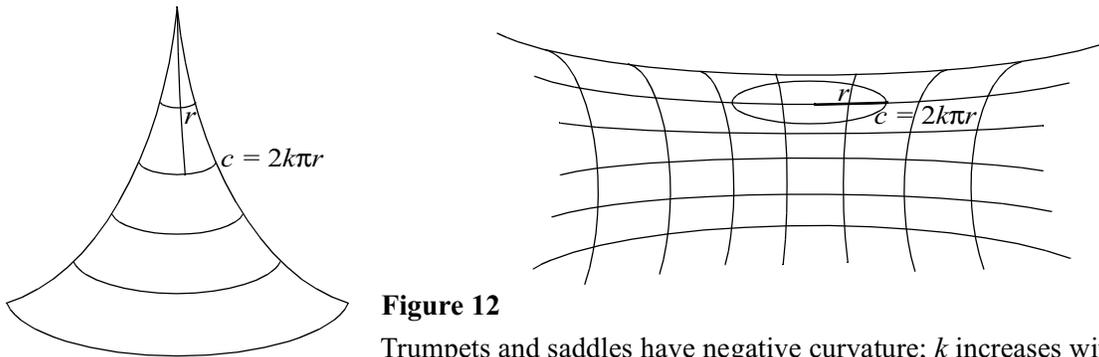

**Figure 12**
Trumpets and saddles have negative curvature; $k$ increases with $r$.

**Singularities.** Imagine a small circle, radius $r$ and centre O. In the region close to O, the geometry may approach a flat geometry, in which the circumference of the circle tends to $2\pi r$ and $k$ tends to 1 as $r$ tends to 0. If $k$ does not get closer and closer to 1, the geometry has a singularity at O.

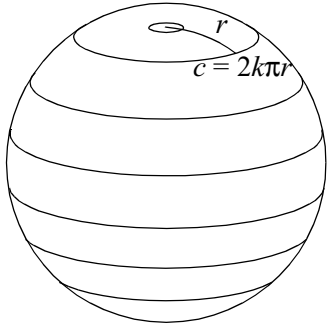

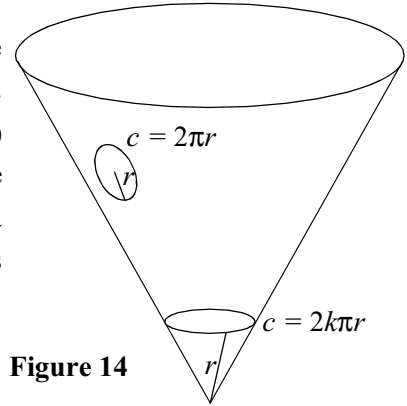

The sphere (figure 13) has positive curvature, since $k \propto c/r$ decreases as $r$ increases. The cone (figure 14) is flat, $k = c/2\pi r$ except at the apex, where the geometry has a cusp, or singularity, $k \to a < 1$ as $r \to 0$.

**Figure 13**

**Figure 14**

## Manifolds

An n-dimensional manifold is a set of n-dimensional coordinate systems such that any point in any coordinate system can be made the origin of another coordinate system in the manifold. The space-time manifold is the set of coordinate systems that can be defined from real or imagined clocks anywhere in the universe by using the radar method. A Riemannian manifold has a metric at each point (according to some authors if the metric is not positive definite it is called semi-Riemannian). A Lorentzian manifold has a Minkowski metric at every point. Space-time is a Lorentzian manifold.

## Tangent Space

A plane can always be found touching a smooth two dimensional surface in three dimensional flat space, such that in a small enough region the geometrical properties of the surface are as close as one likes to those of a corresponding region of the plane. The plane is said to be a tangent to the surface.

An observer at O uses a locally Minkowski reference frame defined by primed axes $\alpha'$ for $\alpha = 0, 1, 2, 3$ (figure 15). The metric is Minkowski at O, and if the observer at the origin of a reference frame does not know, or does not wish to take account of, the metric at all points in space-time may he use Euclidean trigonometry to define a 'distance' between any two points, based only the measurements of distance and direction he can make where he stands. He will then have a space defined in a visible region of the universe such that the physical metric coincides with Minkowski metric at the origin. $g_{\alpha'\beta'}(0) = \eta_{\alpha'\beta'}$. This is a tangent space. Using Minkowski metric an observer does not find that a one metre rule is always one metre long, or that the laws of physics are everywhere the same (in fact bodies will accelerate at different rates depending on position). The 'real' metric is Minkowski at the origin and in a small enough region bounded in time as well as space, the observer cannot detect the difference between tangent space and real space up to the limits of experimental accuracy. For greater distances the scaling distortions between the real space and tangent space will later be identified with gravitational red shift.

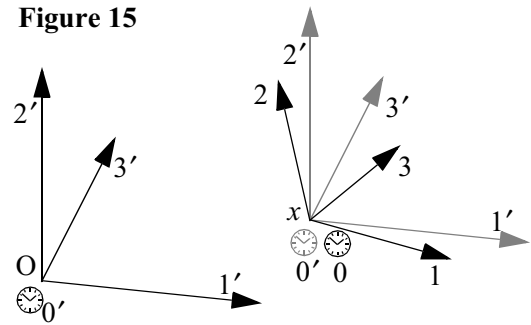

**Figure 15**

It is often convenient to think in terms of tangent space because it is flat, can be drawn on paper and corresponds to the normal Euclidean way in which we most easily think of space. Tangent space has the property of Newtonian absolute space that it is always and everywhere unchanging without relation to anything external. It is not conceived as being "absolute" in a Riemannian manifold, there being a different tangent space at each point. But tangent space is the flat space which we generally use to think of the world about us.

**Coordinate Space**

The intrinsic curvature of space-time can conceived in terms of the stretching or distortion of a map drawn in coordinate space. Such a map is also called a chart. Let coordinate space be denoted by primed axes $\alpha'$ for $\alpha = 0, 1, 2, 3$ with Minkowski metric, $\eta_{\alpha'\beta'}$ everywhere. $\eta_{\alpha'\beta'}$ is not the physical metric, but is an abstract metric used for mapping. We distinguish the scalar product, 3.10, calculated with $g_{\alpha'\beta'}(x)$ from the coordinate product calculated with $\eta_{\alpha'\beta'}$. With a change of scale $\eta_{\alpha'\beta'}$ corresponds to the Euclidean metric of the paper in figure 8 as distinct from the spherical metric of geography. Because coordinate space is identical to itself under changes of scale it provides a "surface" on which to draw charts of (regions of) the universe. The tangent space at $x$ is a coordinate space such that $g_{\alpha'\beta'}(x) = \eta_{\alpha'\beta'}$.

In general straight lines in coordinate space are not geometrically straight, but for a sufficiently short line segment the deviation from straightness is not detectable. A short rod placed at $x$ will appear as a small displacement vector $A^{\alpha'}$ defined as usual as the difference in the coordinates of one end of the rod from the other. A coordinate space vector, designated with a bar to distinguish it from a vector, is defined by

$$\bar{A}^{\alpha'} = k^{\alpha'}_{\beta'}(x) A^{\beta'} \qquad 4.1$$

so that the scalar product of the vectors $A^{\alpha'}$ and $B^{\alpha'}$ defined at $x$ is equal to the coordinate product of the corresponding coordinate space vectors $\bar{A}^{\alpha'}$ and $\bar{B}^{\alpha'}$. I.e.

$$g_{\alpha'\beta'}(x) A^{\alpha'} B^{\beta'} = \eta_{\mu'\nu'} \bar{A}^{\mu'} \bar{B}^{\nu'} = \eta_{\mu'\nu'} k^{\mu'}_{\alpha'}(x) k^{\nu'}_{\beta'}(x) A^{\alpha'} B^{\beta'} \qquad 4.2$$

This is true for any vectors $A^{\alpha'} B^{\alpha'}$ so

$$g_{\alpha'\beta'}(x) = \eta_{\mu'\nu'} k^{\mu'}_{\alpha'}(x) k^{\nu'}_{\beta'}(x) \qquad 4.3$$

4.3 is similar to 3.22 using local Minkowski coordinates with an origin at $x$ but it gives the metric in terms of the variable scale coefficients $k^{\alpha'}_{\beta'}(x)$ of vectors viewed in coordinate space, showing that coordinate space incorporates scaling distortions equivalent to coordinate transformation from a local Minkowski frame at $x$. 4.1 shows that the coefficients of the coordinate space vector are equal to the coefficients of the vector in local Minkowski coordinates.

**Parallel Displacement**

An identical rod is physically placed at $x + dx$, so that its coordinate space vector is parallel to $\bar{A}^{\mu}$. For a small displacement, $dx$, the rod will be described by a vector $B^{\mu}$ at $x + dx$ whose length is unchanged, $A^2 = B^2$. By parallel we mean that coordinate space components are proportional

$$\bar{B}^{\alpha'} = k^{\alpha'}_{\beta'}(x + dx) B^{\beta'} \propto \bar{A}^{\alpha'} \qquad 4.4$$

In general this is not equality, and it is clear from 4.1 that $B^{\mu} \neq A^{\mu}$ if $k^{\alpha'}_{\beta'}(x)$ is not constant. $B^{\mu}$ is the result of parallel displacement of $A^{\mu}$. Because the coefficients of the coordinate space vector are equal to those of the vector in local Minkowski coordinates, 4.4 can be rewritten in local Minkowski coordinates at $x$

$$B^\mu = k^\mu_{\beta'}(x+dx)B^{\beta'} \propto A^\mu \qquad 4.5$$

Thus the effect of parallel displacement in any coordinate space is identical to its effect in the tangent space of an observer at $x$. So parallel displacement is not coordinate dependent, and corresponds to the intuitive notion that two rods can be placed parallel in a small region of space. In section 7, *Differential Geometry*, a formula for change under parallel displacement will be found a without reference to the primed coordinates of a coordinate space.

**Parallel Transport**

In a curved space we cannot give meaning to parallel vectors defined at different points, as one can see by thinking of the curved surfaces described above. However a coordinate space is flat, and we can define parallel in flat space. If we define a path through curved space such that in each region of the path space is near flat then we can define parallel vectors within each region, and so we can define a family of parallel vectors along a path. This family defines the parallel transport of a vector along a path. Parallel transport around a closed loop in general results in a vector at the same point but in a different direction from the original.

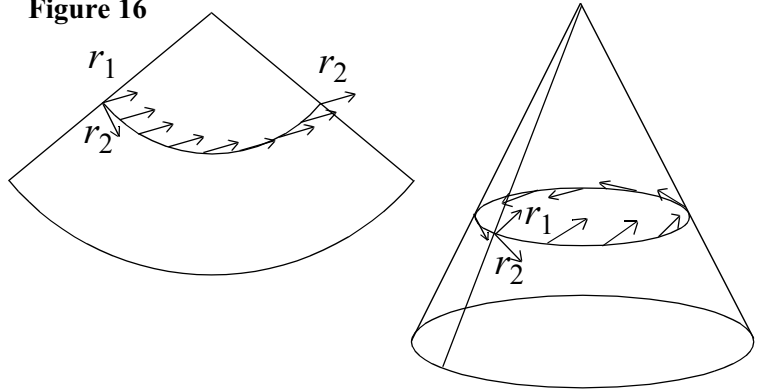

Figure 16

For example since the geometry of a cone (figure 16) is flat except at the apex we can cut it radially to the apex and roll it out flat. Then we can use parallel transport take the vector $r_1$ to $r_2$ along a path about the apex. When the cone is rolled back up $r_2$ does not coincide with $r_1$. In space-time parallel transport will be defined as the cumulative effect of parallel displacement over small distances along a path, as given in 4.4.

**Geodesics**

**Definition:** A geodesic is the path defined by parallel transport of a vector which is always tangential to the path it traces out.

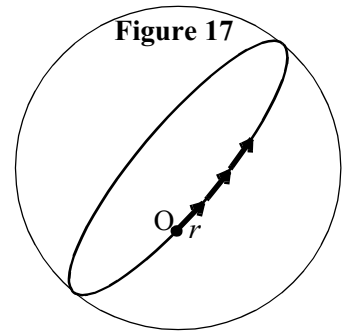

Figure 17

As seen in figure 17, starting from any point O in the surface of a sphere and an infinitesimal vector $r$ in any direction acting at O, we can trace a path in the direction of $r$, and parallel transport $r$ along this path. The result is a great circle, being a geodesic on the sphere (the terminology that $r$ is infinitesimal will mean that a limit is taken such that $r$ is small enough that theoretical errors are less than experimental errors.

## 5 Gravity

**The Principle of Equivalence**

It turns out that in tangent space inertial bodies accelerate at differing rates depending on position, violating Newton's first law. In Newtonian mechanics we may distinguish between force, impressed force and inertial force. A force is anything which causes an acceleration according to the second law. Impressed force implies the action of one body on another. An *inertial* (from latin *not acting*) force is caused not by the action of one body on another but by the choice of reference frame. The coriolis and centrifugal forces in rotating frames, and *g*-forces in accelerated frames are inertial. Because inertial forces are not action, they produce no third law reaction. Because the coriolis and centrifugal forces do not exist in inertial reference frames they are often regarded as fictional, but a general treatment of reference frames incorporates both inertial and non-inertial frames, and inertial forces exist by the second law definition of force. In curved space time Newton's first law is restored by the definition *gravity is the inertial force appearing in an inertial tangent space*. The acceleration of inertial matter due to gravity at the surface of the earth will be understood in an inertial tangent space with an origin at the centre of the earth. Then local reference frames fixed at the surface of the earth accelerate relative to matter in free fall, and hence *gravity is equivalent to the g-force caused by the acceleration of a local reference frame relative to inertial matter.* This is the principle of equivalence

**Gravitational Red Shift**

According to the principle of general relativity the laws of physics in locality are the same as the the laws of physics in any other locality. In each locality the laws of physics are expressed with respect to time determined by an unperturbed or inertial clock in that locality. We have no evidence on which to assume that two clocks at different places can be synchronised even if they are stationary with respect to each other. A second is a second as defined by the behaviour of $Cs^{133}$ in a particular location. It cannot be assumed that this will be the same as a second determined by $Cs^{133}$ in a different place. As with the special theory the relationship between the rate at which the clocks determine the time is measured by red shift. Special relativity studied the consequences of red shift due to difference in speed of clocks, whereas general relativity will study the consequence due to difference in location.

Consider first a static system. Moving clocks may be treated by observing that space local to the clock is Minkowski and applying Lorentz transformation. A space-time diagram (figure 18), is drawn using tangent space, such that light is drawn at $45^o$ and lines of equal time are horizontal. Just as the observer measures locally Minkowski coordinates $x^\mu$ referred to his clock, so locally Minkowski primed coordinates $x^{\mu'}$ can be set up referring to the distant clock. As with the Doppler shift used in the *k*-calculus for special relativity, the time interval between two signals sent from the distant clock to the observer is multiplied by red shift, $k$, $x^0 = x^{0'}/k$ (by considering the signals as the start and stop of a burst of light of a set number of wavelengths of a set frequency). For red shift $k > 1$ the observer's clock is faster, so that, in the tangent space of the distant clock, the coordinate distance between the ends of a rod near the observer is proportionately shorter that the metric distance, or real length of the rod measured locally by the observer from the definition based on light speed. Hence $x^1 = kx^{1'}$. Due to spherical symmetry there is no change in the relationship between metric distance and coordinate distance associated with rotation, so that the general transformation law $x^\mu = k^\mu_{\nu'} x^{\nu'}$ requires

$$k^\mu_{\nu'} = \begin{bmatrix} 1/k & 0 & 0 & 0 \\ 0 & k & 0 & 0 \\ 0 & 0 & 1 & 0 \\ 0 & 0 & 0 & 1 \end{bmatrix} \quad 5.1$$

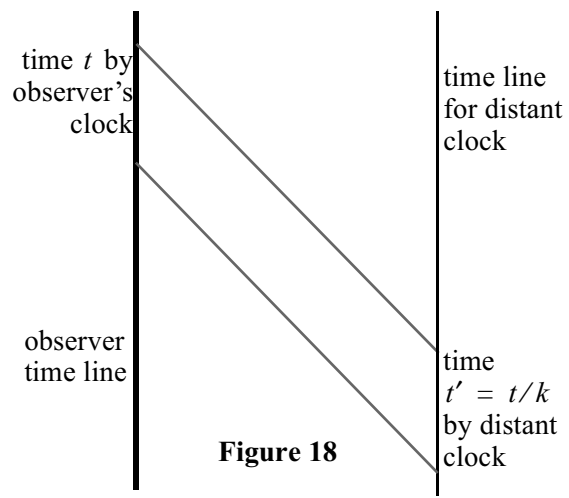

**Figure 18**

So the matrix $k^\mu_{\nu'}$ depends entirely on a single measurable quantity, namely red shift. More generally $k^\mu_{\nu'}$ for a moving observers is found from the static case by applying Lorentz transformation. The net red shift contains both gravitational and Doppler parts.

**Geometric Optics**

Geometric optics is the study of the refraction of the wave function due to the geometry of space-time. The wave function may refer to light or to the quantum mechanical wave function of a body. Red shift is a change in wavelength and causes a change in momentum, much as the change in wavelength of light passing through a medium causes optical refraction. In this intuitive approach the refraction of the wave function will be seen underlying the geodesic motion of classical matter.

Using natural units in which $\hbar = 1$, for an initial wave function $f(x) = \langle x|f\rangle$ at time $t_0$, the momentum space wave function $f(p) = \langle p|f\rangle$ is the transform of the initial wave function

$$\langle p|f\rangle = \left(\frac{1}{2\pi}\right)^{\frac{3}{2}} \int d^3x \langle x|f\rangle e^{-i x \cdot p} \quad 5.2$$

where $p \cdot x$ is defined as in [3.19] but using a metric for three dimensional space at time $t_0$, not Minkowski metric, and the integral is taken over a synchronous slice of a chart containing motion we are interested in (the chart may be bounded and this is approximate to the normal formulation of quantum mechanics on $\mathbb{R}^3$). The initial wave function is the inverse transform

$$\langle x|f\rangle = \left(\frac{1}{2\pi}\right)^{\frac{3}{2}} \int d^3p \langle p|f\rangle e^{i p \cdot x} \quad 5.3$$

Using a metric for curved space $p$ must be treated as a vector field, not a single vector as for quantum mechanics on flat space. Given the value of $p$ at the origin, the field is found by parallel displacement on the synchronous slice $x^0 = t_0$.

**Geodesic Motion**

Define a vector $p = (E, \boldsymbol{p})$ at O so that $E = p^0$ satisfies the mass shell condition 3.2 evaluated with the metric for three dimensional space at time $t_0$. $p$ is determined by parallel displacement along a synchronous slice and for small time increments may be determined by a synchronous parallel displacement in our reference frame, followed by a synchronous parallel displacement in the frame of a moving observer, back to the space origin of our reference frame. For small displacements in locally Minkowski space this is equivalent to a single parallel displacement on the time axis. Repeating the argument incrementally $p = (E, \boldsymbol{p})$ is parallel transported to all points on the manifold. Then the wave function for an inertial body prior to a measurement at time $x^0$

$$\langle x|f \rangle = \left(\frac{1}{2\pi}\right)^{\frac{3}{2}}\int d^3p \, \langle p|f \rangle \, e^{ip \cdot x} \qquad 5.4$$

Thus covariance effectively determines time evolution, as well as conservation of energy and momentum in locally Minkowski regions of space (this is a form of Noether's theorem, and like it depends on homogeneity). For a classical body the wave function is confined in both momentum space and coordinate space (to the limits of experimental accuracy) and the motion of the body in space-time is in the direction of $p$, unaffected by observation. Geodesic motion follows as the cumulative effect of parallel displacement over small time increments.

### A Correction to Radar

A natural modification to the analysis of radar is to hypothesise a time delay between absorption and emission in proper time of a fundamental particle (typically an electron) reflecting electromagnetic radiation. In a coordinate system in which a particle is stationary energy-momentum is $p = (M, 0)$ where $M$ is the rest mass of the particle, and we postulate a proper time delay $4GM$, where $4G$ is a constant of proportionality. This is not the usual assumption of general relativity, but it will be shown that it leads to Schwarzschild geometry and hence to Einstein's field equation.

To affect the geometry of space-time the delay between absorption and emission must be a fundamental property of nature which must be taken into account in addition to speed in determining the shortest time for the transmission of information. In this case the metric is determined by the maximum theoretical speed of information together with the minimum possible time between absorption and emission. This minimum time is fundamental property of the interactions between particles, and can be calculated from knowledge of the gravity due to a single elementary particle (electron or quark). It is not the purpose of this paper to study the truth of this hypothesis, merely to use it as an intuitive assumption leading to Einstein's general theory of relativity, but it is interesting to observe that if proper time delay is real then it relates gravitational mass to inertial mass. If, for example, the interactions of a muon are identical to, but much less frequent than, the interactions of an electron then we would expect to observe a proportionately smaller change in motion from the same stimulus.

### A Curved Space-time Diagram (figure 19)

A single elementary particle at an exact position (*eigenstate*) has spherical symmetry and space-time diagrams may be used to show a radial coordinate without loss of generality. The reflection of a radar pulse is now seen as two events, absorption, A, and emission, E. A and E have the same space-time coordinate, but they are separated by a time $4kGM$ in coordinates in which red shift $k = k(r)$ goes to zero at the position of reflection. Thus the reflection appears instantaneous in a frame defined by radar, but in proper time and in a tangent space at O the time between them is $4GM$.

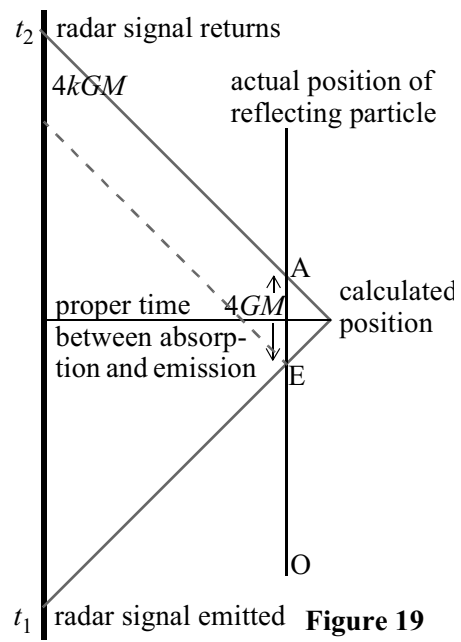

Figure 19

## 6 Schwarzschild

We seek to analyse the gravitational effect of the elementary particle at O in figure 19. Choose a tangent space using primed radial coordinates with the gravitating particle at the origin. The observer uses unprimed radial coordinates with the origin translated to O, so that $k^\mu_{\nu'}$ is given by 5.1. Draw a synchronous slice through the gravitating particle (figure 20) in tangent space and superpose the unprimed coordinates. The observer's radial coordinate is stretched so that the origins coincide and the particle is shown as a point in the diagram. The stretch is given by red shift $k$,

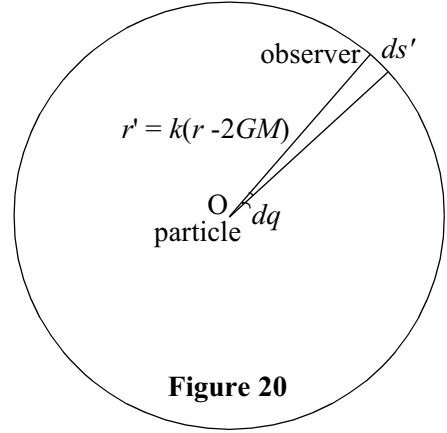

Figure 20

$$r' = k(r - 2GM) \qquad 6.1$$

Consider an infinitesimal coordinate length $ds'$ along the circumference.

$$ds' = r'd\theta = k(r - 2GM)d\theta \qquad 6.2$$

using 6.1. For red shift, $k > 1$, the observer's clock runs fast, so distances measured by proper time along the circumference are correspondingly shorter. Hence in the unprimed coordinates

$$ds' = rd\theta/k \qquad 6.3$$

Comparing 6.3 and 6.2

$$k^{-2} = 1 - \frac{2GM}{r} \qquad 6.4$$

From 4.3, 5.1 and 6.4 we find the familiar form of the Schwarzschild metric,

$$g_{\alpha'\beta'} = \eta_{\mu\nu} k^\mu_{\alpha'} k^\nu_{\beta'} = \begin{bmatrix} 1 - \frac{2GM}{r} & 0 & 0 & 0 \\ 0 & -\left(1 - \frac{2GM}{r}\right)^{-1} & 0 & 0 \\ 0 & 0 & -r^2 & 0 \\ 0 & 0 & 0 & -r^2\sin^2\theta \end{bmatrix} \qquad 6.5$$

**The Newtonian Approximation**

Red shift affects the frequency of the wave function, and hence the energy of a particle, according to the relation

$$k^0_{0'} p^{0'} = \text{const} \qquad 6.6$$

So the classical energy $E$ satisfies

$$E = \langle P^0 \rangle \propto 1/k^0_{0'} = k = \left(1 - \frac{2GM}{r}\right)^{-1/2} \approx 1 + \frac{GM}{r} \qquad 6.7$$

For a body of mass $m$ the constant of proportionality is fixed at $r = \infty$

$$E = m + \frac{GMm}{r} \qquad 6.8$$

showing the gravitational potential in the Newtonian approximation

# 7 Differential Geometry

In differential geometry calculus is used to precisely describe all possibilities for non-Euclidean geometry. This section is included to show that the treatment of parallel displacement is equivalent to that used in other accounts, to establish Einstein's field equation, and for students intending further study.

**Partial Differentiation**

Given any quantity $A = A(x)$ which is a function of position, we can differentiate with respect to each coordinate $x^\mu$. This is the partial derivative, or ordinary derivative, denoted by

$$A_{,\mu} = \partial_\mu A = \frac{\partial A}{\partial x^\mu} \qquad 7.1$$

By convention only one comma is used for a second partial derivative

$$A_{,\mu\nu} = A_{,\mu,\nu} \qquad 7.2$$

For a vector $x^{\mu'}$ in primed coordinates

$$x^{\mu'} = \frac{\partial x^{\mu'}}{\partial x^\nu} x^\nu = x^{\mu'}_{,\nu} x^\nu \qquad 7.3$$

Comparison with the vector transformation law, 3.5, shows

$$k^{\mu'}_\nu = x^{\mu'}_{,\nu} \qquad 7.4$$

**Corollary:** Since the order of partial differentiation makes no difference

$$k^{\mu'}_{\nu,\sigma} = x^{\mu'}_{,\nu\sigma} = k^{\mu'}_{\sigma,\nu} \qquad 7.5$$

**Theorem:** The partial derivative of a scalar field is a covariant vector field

*Proof:* The vector transformation law follows from the laws of partial differentiation:

$$\frac{\partial A}{\partial x^\nu} = \frac{\partial x^{\mu'}}{\partial x^\nu} \frac{\partial A}{\partial x^{\mu'}} = k^{\mu'}_\nu \frac{\partial A}{\partial x^{\mu'}} \qquad 7.6$$

The partial derivative of a vector field is not a tensor because the definition of a derivative requires

$$\frac{\partial A}{\partial x^\nu} = \lim_{\delta x^\nu \to 0} \frac{A(x+\delta x) - A(x)}{\delta x^\nu} \qquad 7.7$$

and if $A$ is a vector field then $A(x)$ and $A(x+\delta x)$ are vectors at different points in space-time, and the subtraction between them is not legitimate. Subtraction of vectors at different points can be defined using parallel displacement in coordinate space.

**Parallel Displacement**

**Lemma:** Let $D^\mu$ be a contravariant vector acting at $x$. Let $C_\mu = D^{\alpha'} k_{\alpha'\mu}(x)$, where $\alpha'$ refer to coordinate space. Then $C^2 = D^2$

*Proof:*
$$C^2 = g^{\mu\nu} D^{\alpha'} k_{\alpha'\mu}(x) D^{\beta'} k_{\beta'\nu}(x)$$
$$= g_{\mu\nu} D^{\alpha'} k^\mu_{\alpha'}(x) D^{\beta'} k^\nu_{\beta'}(x) \qquad 7.8$$
$$= g_{\alpha'\beta'}(x) D^{\alpha'} D^{\beta'} = D^2 \qquad \text{using 3.22}$$

By the lemma multiplying both sides of 4.5 by $k_{\alpha'\mu}(x+dx)$ gives equality in magnitude, since $A$ and

$B$ represent identical rods with $A^2 = B^2$. Since they are also proportional we have equality

$$A^{\alpha'}k_{\alpha'\mu}(x+dx) = B^{\nu}k_{\nu}^{\alpha'}(x+dx)k_{\alpha'\mu}(x+dx)$$
$$= B^{\nu}g_{\mu\nu}(x+dx) = B_{\mu}(x+dx) \quad 7.9$$

Hence to first order in $dx$

$$B_{\mu}(x+dx) = A^{\alpha'}[k_{\alpha'\mu}(x) + k_{\alpha'\mu,\sigma}dx^{\sigma}] \quad 7.10$$
$$= A^{\nu}k_{\nu}^{\alpha'}(x)[k_{\alpha'\mu} + k_{\alpha'\mu,\sigma}dx^{\sigma}] \quad 7.11$$
$$= A_{\mu} + A^{\nu}k_{\nu}^{\alpha'}k_{\alpha'\mu,\sigma}dx^{\sigma} \quad 7.12$$

$B_{\mu}$ is described as the result of parallel displacement of $A_{\mu}$ to $x+dx$. The change in $A_{\mu}$ under parallel displacement to $x+dx$ is

$$dA_{\mu} = B_{\mu} - A_{\mu} = A^{\nu}k_{\nu}^{\alpha'}k_{\alpha'\mu,\sigma}dx^{\sigma} \quad 7.13$$

**Christoffel Symbols**

**Definition:** A Christoffel symbol of the first kind is

$$\Gamma_{\mu\nu\sigma} = (g_{\mu\nu,\sigma} + g_{\mu\sigma,\nu} - g_{\nu\sigma,\mu})/2 \quad 7.14$$

Referred to a primed Minkowski tangent space an origin at $x$ $g_{\mu\nu} = \eta_{\alpha'\beta'}k_{\mu}^{\alpha'}k_{\nu}^{\beta'}$ (from 4.3). Differentiating

$$g_{\mu\nu,\sigma} = \eta_{\alpha'\beta'}(k_{\mu,\sigma}^{\alpha'}k_{\nu}^{\beta'} + k_{\mu}^{\alpha'}k_{\nu,\sigma}^{\beta'})$$

Hence

$$\Gamma_{\mu\nu\sigma} = \eta_{\alpha'\beta'}(k_{\mu,\sigma}^{\alpha'}k_{\nu}^{\beta'} + k_{\mu}^{\alpha'}k_{\nu,\sigma}^{\beta'} + k_{\mu,\nu}^{\alpha'}k_{\sigma}^{\beta'} + k_{\mu}^{\alpha'}k_{\sigma,\nu}^{\beta'} - k_{\nu,\mu}^{\alpha'}k_{\sigma}^{\beta'} - k_{\nu}^{\alpha'}k_{\sigma,\mu}^{\beta'})/2 \quad 7.15$$

$\eta_{\alpha'\beta'}$ is symmetrical in the indices so $\alpha'$ and $\beta'$ can be interchanged in the last three terms. Then using 7.5 this reduces to

$$\Gamma_{\mu\nu\sigma} = g_{\alpha'\beta'}k_{\mu}^{\alpha'}k_{\nu,\sigma}^{\beta'} = k_{\mu}^{\alpha'}k_{\alpha'\nu,\sigma} \quad 7.16$$

Then, from 7.13, the change in $A_{\mu}$ under parallel displacement can be expressed in terms of the physical metric, without reference to the primed coordinates of a tangent space

$$dA_{\nu} = \Gamma_{\mu\nu\sigma}A^{\mu}dx^{\sigma} \quad 7.17$$

**Definition:** A Christoffel symbol of the second kind is defined by raising the first index

$$\Gamma_{\nu\sigma}^{\mu} = g^{\mu\lambda}\Gamma_{\lambda\nu\sigma} \quad 7.18$$

Then standard form for the change in covariant components under parallel displacement is

$$dA_{\nu} = \Gamma_{\nu\sigma}^{\mu}A_{\mu}dx^{\sigma} \quad 7.19$$

Given a second vector $B^{\nu}$, $A_{\nu}B^{\nu}$ is a scalar, and so unchanged under parallel displacement

$$0 = d(A_{\nu}B^{\nu}) = A_{\nu}dB^{\nu} + dA_{\nu}B^{\nu} = A_{\nu}dB^{\nu} + A_{\mu}\Gamma_{\nu\sigma}^{\mu}dx^{\sigma}B^{\nu} \quad 7.20$$

7.20 holds for any $A_{\mu}$, so the change in of contravariant components under parallel displacement is

$$dB^{\nu} = -\Gamma_{\mu\sigma}^{\nu}B^{\mu}dx^{\sigma} \quad 7.21$$

**Covariant Differentiation**

The covariant derivative of a contravariant vector $A^\mu$ is defined from the ordinary definition of derivative 7.7 by subtracting the parallel displacement $A^\mu(x) + dA^\nu$ of $A^\mu(x)$ at $x + \delta x$, using 7.21

$$A^\mu_{;\nu}(x) = \nabla_\nu A^\mu = \frac{dA^\mu}{dx^\nu} = \lim_{\delta x^\nu \to 0} \frac{A^\mu(x + \delta x) - A^\mu(x) - dA^\nu}{\delta x^\nu} \qquad 7.22$$

$$A^\mu_{;\nu} = A^\mu_{,\nu} + \Gamma^\mu_{\lambda\nu} A^\lambda \qquad 7.23$$

The covariant derivative of a covariant vector $A_\mu$ is found in the same way using 7.19

$$A_{\mu;\nu} = A_{\mu,\nu} - \Gamma^\lambda_{\mu\nu} A_\lambda \qquad 7.24$$

The covariant derivative of a scalar field is simply the partial derivative, as scalars are invariant under displacement. The covariant derivative of a vector field is a tensor field, because the index μ transforms as a vector, and the theorem for partial differentiation of scalar can be applied to the components. It is a straightforward exercise to verify this directly by a coordinate transformation, and also to show that covariant differentiation of the tensor product $A_\mu B^\nu$ satisfies the Leibniz rule

$$(A_\mu B^\nu)_{;\sigma} = B^\nu A_{\mu;\sigma} + A_\mu B^\nu_{;\sigma} \qquad 7.25$$

and that

$$(A_\mu B^\nu)_{;\sigma} = (A_\mu B^\nu)_{,\sigma} - \Gamma^\lambda_{\mu\sigma} A_\lambda B^\nu + \Gamma^\nu_{\lambda\sigma} A_\mu B^\lambda \qquad 7.26$$

And, since any tensor is a sum of tensor products of vectors, for a tensor $C^\nu_\mu$ with two indices

$$C^\nu_{\mu;\sigma} = C^\nu_{\mu,\sigma} - \Gamma^\lambda_{\mu\sigma} C^\nu_\lambda + \Gamma^\nu_{\lambda\sigma} C^\lambda_\mu \qquad 7.27$$

and more generally we can form the covariant derivative of a tensor with any number of indices, top an bottom by adding a term for each index. In particular, for the metric tensor

$$g_{\mu\nu;\sigma} = g_{\mu\nu,\sigma} - \Gamma^\lambda_{\mu\sigma} g_{\lambda\nu} - \Gamma^\lambda_{\nu\sigma} g_{\mu\lambda} = g_{\mu\nu,\sigma} - \Gamma_{\nu\mu\sigma} - \Gamma_{\mu\nu\sigma} = 0 \qquad 7.28$$

by direct application of 7.14. So the metric tensor behaves as a constant with respect to covariant differentiation.

**The Curvature Tensor**

By convention only one semicolon is used for a second covariant derivative; for any tensor field $A$

$$A_{;\mu\nu} = A_{;\mu;\nu} \qquad 7.29$$

Since covariant differentiation of a scalar field $A$ is partial differentiation

$$A_{;\mu\nu} = A_{,\mu;\nu} = A_{,\mu\nu} - \Gamma^\lambda_{\mu\nu} A_{,\lambda} \qquad 7.30$$

and since the Christoffel symbol 7.14 is symmetric under interchange of the last two indices, the order of covariant differentiation of a scalar makes no difference. However for a vector field $A_\nu$

$$A_{\nu;\rho\sigma} = A_{\nu;\rho,\sigma} - \Gamma^\lambda_{\nu\sigma} A_{\lambda;\rho} - \Gamma^\lambda_{\rho\sigma} A_{\nu;\lambda} \qquad 7.31$$

$$= (A_{\nu,\rho} - \Gamma^\lambda_{\nu\rho} A_\lambda)_{,\sigma} - \Gamma^\lambda_{\nu\sigma}(A_{\lambda,\rho} - \Gamma^\kappa_{\lambda\rho} A_\kappa) - \Gamma^\lambda_{\rho\sigma}(A_{\nu,\lambda} - \Gamma^\kappa_{\nu\lambda} A_\kappa) \qquad 7.32$$

$$\begin{aligned} = A_{\nu,\rho\sigma} - \Gamma^\lambda_{\nu\rho} A_{\lambda,\sigma} - \Gamma^\lambda_{\nu\sigma} A_{\lambda,\rho} - \Gamma^\lambda_{\rho\sigma} A_{\nu,\lambda} \\ - (\Gamma^\kappa_{\nu\rho,\sigma} + \Gamma^\lambda_{\nu\sigma} \Gamma^\kappa_{\lambda\rho} + \Gamma^\lambda_{\rho\sigma} \Gamma^\kappa_{\nu\lambda}) A_\kappa \end{aligned} \qquad 7.33$$

Then interchange the order of ρ and σ and subtract from the previous express

$$A_{\nu;\rho\sigma} - A_{\nu;\sigma\rho} = R^{\kappa}_{\nu\rho\sigma} A_{\kappa} \qquad 7.34$$

where

$$R^{\kappa}_{\nu\rho\sigma} = \Gamma^{\kappa}_{\nu\sigma,\rho} - \Gamma^{\kappa}_{\nu\rho,\sigma} + \Gamma^{\lambda}_{\nu\sigma}\Gamma^{\kappa}_{\lambda\rho} - \Gamma^{\lambda}_{\nu\rho}\Gamma^{\kappa}_{\lambda\sigma} \qquad 7.35$$

Since the left hand side of 7.34 transforms as a tensor, the right hand side is a tensor, and since this true for any vector $A_\nu$ $R^{\kappa}_{\nu\rho\sigma}$ is a tensor. $R^{\kappa}_{\nu\rho\sigma}$ is the Riemann Christoffel tensor, or curvature tensor. A simplified measure of curvature, the Ricci tensor, is found by contracting the first and last indices

$$R_{\nu\rho} = R^{\kappa}_{\nu\rho\kappa} \qquad 7.36$$

And a scalar curvature is found by contracting with $g^{\nu\rho}$

$$R = g^{\nu\rho} R_{\nu\rho} \qquad 7.37$$

**Einstein's Field Equation**

In the static system shown in figure 18 there may be two causes of gravitational red shift; $k$ may be a function of position or distance in empty space, and $k$ may be directly dependent on the distribution of matter. In either case a tensor equation is required to describe the geometry, and it is convenient to use an equation for curvature. In the vacuum case $k$ is a function of distance, and from the homogeneity of the vacuum it follows that this is described by constant curvature, which gives the cosmological constant term in Einstein's Field equation for empty space,

$$R_{\mu\nu} = \lambda g_{\mu\nu} \qquad 7.38$$

Ignoring this term it is possible to follow the standard argument that the Einstein tensor is proportional to stress energy

$$G^{\alpha\beta} = R^{\alpha\beta} - \tfrac{1}{2}g^{\alpha\beta}R = 8\pi G T^{\alpha\beta} \qquad 7.39$$

but a purpose of this paper is to see that Einstein's field equation 7.39 follows from the definition of a metric defined by two way light speed, after taking into account a small delay in reflection. Since Schwarzschild is a known solution of Einstein's field equation it is sufficient to have shown that the Schwarzschild geometry obtains for a single point particle in an eigenstate of position. Then the operator form of 7.39 holds for one particle, and so it holds generally by linearity of operators in quantum mechanics.

A fundamental particle with an exact position would be a naked singularity, but in practice there is always uncertainty in position, so that the eigenstate $|x\rangle$ of exact position is replaced with the state $|f\rangle$ for which the probability amplitude for finding the particle at a particular position is $\langle x|f \rangle$. For a scalar particle the corresponding energy density operator is $P^0 = -i\nabla^0$ and the energy density is $\rho = \langle f|x\rangle P^0 \langle x|f\rangle$. In a near eigenstate with the particle at rest we have $\langle P^i \rangle = 0$ for $i = 1, 2, 3$ and scalar curvature

$$R \propto \langle f|x\rangle G P^0 \langle x|f\rangle = G\rho \qquad 7.40$$

So Einstein's Field Equation is satisfied in the rest frame. The general form of the field equation is found by composing a second rank tensor equation which reduces to energy in the rest frame of a par-ticle. For a Dirac particles this is satisfied by

$$T_D^{\alpha\beta} = -i\langle f|\boldsymbol{x}\rangle\gamma^{\alpha}\nabla^{\beta}\langle\boldsymbol{x}|f\rangle: \quad\quad 7.41$$

When a photon transmits energy from one position to another curvature is transmitted with it. The natural conclusion is that electromagnetic energy also generates curvature.

$$T_{EM}^{\alpha\beta} = -i\nabla^{\beta}A^{\alpha}(x) \quad\quad 7.42$$

The full stress energy tensor is the sum of terms 7.41 7.42 for each elementary particle, so Einstein's field equation takes the form

$$G^{\alpha\beta} = 8\pi G(T_D^{\alpha\beta} + T_{EM}^{\alpha\beta}) = 8\pi G T^{\alpha\beta} \quad\quad 7.43$$

## Acknowledgements

I should like to thank a number of physicists who have discussed the content and ideas of this paper, particularly John Farina, Mike Mowbray, Frank Wappler, Eric Forgy, Chris Hillman and John Baez and the moderators of sci.physics.research (John Baez, Matt McIrvin, Ted Bunn & Philip Helbig) for their vigilance in pointing out lack of clarity in expression.

## References

[1]   **Bondi H.:** *Assumption and Myth in Physical Theory*, Cambridge University Press (1967)
[2]   **Francis C. E. H.:** *A Metric from Photon Exchange*, physics/0108012